\documentclass[runningheads]{llncs}

\usepackage{graphicx}
\usepackage{comment}
\usepackage{amsmath,amssymb}
\usepackage{color}

\usepackage[american]{babel}
\usepackage{hyperref}
\addto\extrasamerican{
}

\usepackage{siunitx}

\usepackage{booktabs}
\usepackage{tabularx}
\usepackage[para]{threeparttable}
\usepackage{multirow}
\newcommand{\ra}[1]{\renewcommand{\arraystretch}{#1}}

\newif\ifreview
\reviewfalse

\ifreview
	\usepackage{lineno}

	\linenumbers
\fi

\graphicspath{{./images/}}

\begin{document}


\def\SubNumber{000}

\def\GCPRTrack{Regular Track}

\ifreview
\else
\title{AudioCLIP: Extending CLIP to Image, Text and Audio\thanks{This work was supported by the BMBF projects ExplAINN (Grant 01IS19074), EDeL (Grant 01IS19075) and the TU Kaiserslautern PhD program.}}
\fi

\ifreview
\else

	\author{
	    Andrey Guzhov\inst{1,2}  
	    \and
	    Federico Raue\inst{1}  
	    \and
	    J\"orn Hees\inst{1}  
	    \and
	    Andreas Dengel\inst{1,2}  
	}
	
	\authorrunning{A. Guzhov et al.}

	\institute{
	    DFKI GmbH, Trippstadter Str. 122, 67663 Kaiserslautern, Germany\\
	    \and
	    TU Kaiserslautern, Kaiserslautern, Germany\\
	    \email{firstname.lastname@dfki.de}
	}
\fi

\maketitle              

\begin{abstract}
In the past, the rapidly evolving field of sound classification greatly benefited from the application of methods from other domains.
Today, we observe the trend to fuse domain-specific tasks and approaches together, which provides the community with new outstanding models.

In this work, we present an extension of the \mbox{CLIP} model that handles audio in addition to text and images.
Our proposed model incorporates the \mbox{ESResNeXt} audio-model into the \mbox{CLIP} framework using the \mbox{AudioSet} dataset.
Such a combination enables the proposed model to perform bimodal and unimodal classification and querying, while keeping \mbox{CLIP's} ability to generalize to unseen datasets in a zero-shot inference fashion.

AudioCLIP achieves new state-of-the-art results in the Environmental Sound Classification (ESC) task, out-performing other approaches by reaching accuracies of $90.07\,\%$ on the \mbox{UrbanSound8K} and $97.15\,\%$ on the \mbox{ESC-50} datasets.
Further it sets new baselines in the zero-shot ESC-task on the same datasets ($68.78\,\%$ and $69.40\,\%$, respectively).

Finally, we also assess the cross-modal querying performance of the proposed model as well as the influence of full and partial training on the results.
For the sake of reproducibility, our code is published.

\keywords{Multimodal learning \and Audio classification  \and Zero-shot inference.}
\end{abstract}

\section{Introduction} \label{sec:intro}
The latest advances of the sound classification community provided many powerful audio-domain models that demonstrated impressive results.
Combination of widely known datasets -- such as \mbox{AudioSet} \cite{gemmeke2017audioset}, \mbox{UrbanSound8K} \cite{salamon2014us8k} and \mbox{ESC-50} \cite{piczak2015esc} -- and domain-specific and inter-domain techniques conditioned the rapid development of audio-dedicated methods and approaches \cite{kumar2020weanet,guzhov2021esrnx,verbitskiy2021erann}.

Previously, researchers were focusing mostly on the classification task using the audible modality exclusively.
In the last years, however, popularity of multimodal approaches in application to audio-related tasks has been increasing \cite{kim2019audiocaps,alayrac2020versatile,zhang2021sscl}.
Being applied to audio-specific tasks, this implied the use of either textual or visual modalities in addition to sound.
While the use of an additional modality together with audio is not rare, combination of more than two modalities is still uncommon in the audio domain.
However, the restricted amount of qualitatively labeled data is constraining the development of the field in both, uni- and multimodal directions.
Such a lack of data has challenged the research and sparked a cautious growth of interest for zero- and few-shot learning approaches based on contrastive learning methods that rely on textual descriptions \cite{islam2019soundsemantics,xie2019zero,xie2021zero}.

In our work, we propose an approach to combine a high-performance audio model -- \mbox{ESResNeXt} \cite{guzhov2021esrnx} -- into a contrastive text-image model, namely \mbox{CLIP} \cite{radford2021clip}, thus, obtaining a \emph{tri-modal} hybrid architecture.
The base \mbox{CLIP} model demonstrates impressive performance and strong domain adaptation capabilities that are referred as ``zero-shot inference" in the original paper \cite{radford2021clip}.
To keep consistency with the \mbox{CLIP} terminology, we use the term ``zero-shot" in the sense defined in \cite{radford2021clip}.

As we will see, the joint use of three modalities during the training results in out-performance of previous models in environmental sound classification task, extends zero-shot capabilities of the base architecture to the audio modality and introduces an ability to perform cross-modal querying using text, image and audio in any combination.

The remainder of this paper is organized as follows.
In \autoref{sec:rel_work} we discuss the current approaches to handle audio in a standalone manner as well as jointly with additional modalities.
Then, we describe models that serve as a base of our proposed hybrid architecture in \autoref{sec:model}, its training and evaluation in \autoref{sec:exp_setup} and the obtained results in \autoref{sec:results}.
Finally, we summarize our work and highlight follow-up research directions in \autoref{sec:conclusion}.

\section{Related Work} \label{sec:rel_work}
In this section, we provide an overview of the audio-related tasks and approaches that are intersecting in our work.
Beginning with description of the environmental sound classification task, we connect it to the zero-shot classification through the description of existing methods to handle multiple modalities in a single model.

The environmental sound classification task implies an assignment of correct labels given samples belonging to sound classes that surround us in the everyday life (e.g., ``alarm clock", ``car horn", ``jackhammer", ``mouse clicking", ``cat").
To successfully solve this task, different approaches were proposed that included the use of one- \cite{tokozume2017envnet,tokozume2017envnetv2} or two-dimensional Convolutional Neural Networks (CNN) operating on static \cite{piczak2015cnn,salamon2017cnn,xie2019zero,guzhov2020esrn,kumar2020weanet,palanisamy2020densenet,xie2021zero,gong2021ast,verbitskiy2021erann} or trainable \cite{sailor2017convrbm,guzhov2021esrnx} time-frequency transformation of raw audio.
While the first approaches relied on the task-specific design of models, the latter results confirmed that the use of domain adaptation from visual domain is beneficial \cite{guzhov2020esrn,palanisamy2020densenet,guzhov2021esrnx}.
However, the visual modality was used in a sequential way, implying the processing of only one modality simultaneously.

The joint use of several modalities occurred first in video-related \cite{kim2019audiocaps,zhang2021sscl,dzabraev2021mdmmt} tasks and was adapted to the sound classification task later \cite{islam2019soundsemantics,wang2021multimodal}.
However, despite the multimodal design, such approaches utilized two modalities simultaneously at most, while recent studies suggest that the use of more modalities is beneficial \cite{alayrac2020versatile,akbari2021vatt}.

The multimodal approaches described above share a common key idea of contrastive learning.
Such a technique belongs to the branch of self-supervised learning that, among other features, helps to overcome the lack of qualitatively labeled data.
That makes it possible to apply contrastive learning-based training to the zero-shot classification tasks \cite{islam2019soundsemantics,xie2019zero,xie2021zero}.

Summarizing, our proposed model employs contrastive learning to perform training on textual, visual and audible modalities, is able to perform modality-specific classification or, more general, querying and is implicitly enabled to generalize to previously unseen datasets in a zero-shot inference setup.

\section{Model} \label{sec:model}
\begin{figure}
\includegraphics[trim={2.2cm 21.0cm 1.3cm 2.5cm},clip,width=\textwidth]{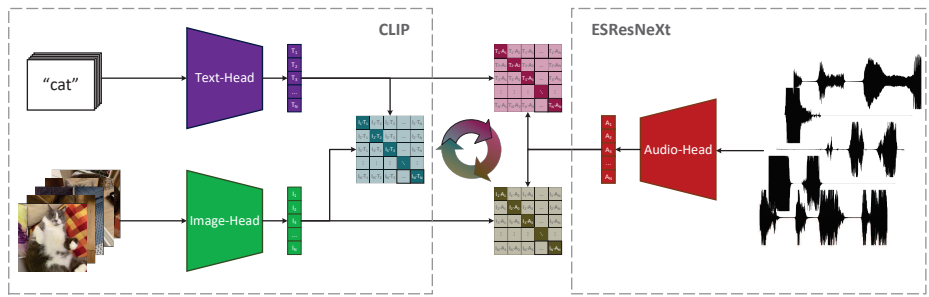}
\caption{
Overview of the proposed \mbox{AudioCLIP} model.
On the left, the workflow of the text-image-model \mbox{CLIP} is shown.
Performing joint training of the text- and image-heads, \mbox{CLIP} learns to align representations of the same concept in a shared multimodal embedding space.
On the right, the audio-model \mbox{ESResNeXT} is shown.
Here, the added audible modality interacts with two others, enabling the model to handle 3 modalities simultaneously.
}
\label{fig:aclp}
\end{figure}

In this section, we describe the key components that make up the proposed model and the way how it handles its input.
On a high level, our hybrid architecture combines a ResNet-based CLIP model \cite{radford2021clip} for visual and textual modalities and an ESResNeXt model \cite{guzhov2021esrnx} for audible modality, as can be seen in \autoref{fig:aclp}.

\subsection{CLIP} \label{sec:model:clip}
Conceptually, the CLIP model consists of two subnetworks: text and image encoding heads.
Both parts of the CLIP model were pre-trained jointly under natural language supervision \cite{radford2021clip}.
Such a training setup enabled the model to generalize the classification ability to image samples that belonged to previously unseen datasets according to the provided labels without any additional fine-tuning.

For the text encoding part, a slightly modified \cite{radford2019transformer} Transformer \cite{vaswani2017attention} architecture was used \cite{radford2021clip}.
For the chosen 12-layer model, the input text was represented by a lower-case byte pair encoding with a vocabulary of size $49\,408$ \cite{radford2021clip}.
Due to computational constraints, the maximum sequence length was clipped at 76 \cite{radford2021clip}.

For the image encoding part of the \mbox{CLIP} model, two different architectures were considered.
One was a Vision Transformer (ViT) \cite{radford2021clip,dosovitskiy2020vit}, whose architecture made it similar to the text-head.
Another option was represented by a modified \mbox{ResNet-50} \cite{he2016resnet}, whose global average pooling layer was replaced by a QKV-attention layer \cite{radford2021clip}.
As we mentioned in \autoref{sec:model:clip}, for the proposed hybrid model we chose the ResNet-based variant of the CLIP model because of its lower computational complexity, in comparison to the ViT-based one.

Given an input batch (text-image pairs) of size $N$, both CLIP-subnetworks produce the corresponding embeddings that are mapped linearly into a multimodal embedding space of size $1\,024$ \cite{radford2021clip}.
In a such setup, CLIP learns to maximize the cosine similarity between matching textual and visual representations, while minimizing it between incorrect ones, which is achieved using symmetric cross entropy loss over similarity measures \cite{radford2021clip}.

\subsection{ESResNeXt} \label{sec:model:esrnx}
For the audio encoding part, we decided to apply \mbox{ESResNeXt} model \cite{guzhov2021esrnx} that is based on the \mbox{ResNeXt-50} \cite{chollet2017resnext} architecture and consists of a trainable time-frequency transformation based on complex frequency B-spline wavelets \cite{teolis1998fbsp}.
The chosen model contains moderate number of parameters to learn (\mbox{$\sim 30$ M}), while performing competitive on a large-scale audio dataset, namely \mbox{AudioSet} \cite{gemmeke2017audioset}, and providing state-of-the-art-level classification results on the \mbox{UrbanSound8K} \cite{salamon2014us8k} and \mbox{ESC-50} \cite{piczak2015esc} datasets.
Additionally, the ESResNeXt model supports an implicit processing of a multi-channel audio input and provides improved robustness against additive white Gaussian noise and sample rate decrease \cite{guzhov2021esrnx}.

\subsection{Hybrid Model -- AudioCLIP} \label{sec:model:audioclip}

In this work, we introduce an additional -- audible -- modality into the novel \mbox{CLIP} framework, which is naturally extending the existing model.
We consider the newly added modality as equally important as the originally present.
Such a modification became possible through the use of the \mbox{AudioSet} \cite{gemmeke2017audioset} dataset that we found suitable for this, as described in \autoref{sec:exp_setup:datasets}.

Thus, the proposed \mbox{AudioCLIP} model incorporates three subnetworks: text-, image- and audio-heads.
In addition to the existing text-to-image-similarity loss term, there are two new ones introduced: text-to-audio and image-to-audio.
The proposed model is able to process all three modalities simultaneously, as well as any pair of them.

\section{Experimental Setup} \label{sec:exp_setup}
In this section, we describe datasets that were used, data augmentation methods we applied, the training process and its corresponding hyper-parameters, finalizing with the performance evaluation methods.

\subsection{Datasets} \label{sec:exp_setup:datasets}
In this work, five image, audio and mixed datasets were used directly and indirectly.
Here, we describe these datasets and define their roles in the training and evaluation processes.

\emph{Composite CLIP Dataset:} In order to train \mbox{CLIP}, a new dataset was constructed by its authors.
It consisted of roughly $400\:\si{\mega}$ text-image pairs based on a set of $\sim 500\:\si{\kilo}$ text-based queries, and each query covered at least $\sim 20\:\si{\kilo}$ pairs \cite{radford2021clip}.
In this work, the \mbox{CLIP} dataset was used indirectly as a weight initializer of the text- and image-heads (\mbox{CLIP} model).

\emph{ImageNet:} \mbox{ImageNet} is a large-scale visual datasets described in \cite{deng2009imagenet} that contains more than \mbox{$1$\:\si{\mega}} images across $1\,000$ classes.
For the purposes of this work, the \mbox{ImageNet} dataset served as a weight initializer of the \mbox{ESResNeXt} model and as a target for the zero-shot inference task.

\emph{AudioSet:} Being proposed in \cite{gemmeke2017audioset}, the \mbox{AudioSet} dataset provides a large-scale collection (\mbox{$\sim 1.8$\:\si{\mega}} \& \mbox{$\sim 20$\:\si{\kilo} evaluation set}) of audible data organized into $527$ classes in a non-exclusive way.
Each sample is a snippet up to $10\:\si{\second}$ long from a {YouTube}-video, defined by the corresponding ID and timings.

For this work, we acquired video frames in addition to audio tracks.
Thus, the \mbox{AudioSet} dataset became the glue between the vanilla CLIP framework and our tri-modal extension on top of it.
In particular, audio tracks and the respective class labels were used to perform image-to-audio transfer learning for the \mbox{ESResNeXt} model, and then, the extracted frames in addition to audio and class names served as an input for the hybrid \mbox{AudioCLIP} model.

During the training part, ten equally distant frames were extracted from a video recording, and one of them was picked randomly ($\sim \mathcal{U}$) and passed through the \mbox{AudioCLIP} model.
In the evaluation phase, the same extraction procedure was performed, with the difference that only a central frame was presented to the model.
Performance metrics are reported based on the evaluation set of the \mbox{AudioSet} dataset.

\emph{UrbanSound8K:} The \mbox{UrbanSound8K} dataset provides $8\,732$ mono- and binaural audio tracks sampled at frequencies in the range $16-48\:\si{\kilo \hertz}$, each track is not longer than $4\:\si{\second}$.
The audio recordings are organized into ten classes: ``air conditioner", ``car horn", ``children playing", ``dog bark", ``drilling", ``engine idling", ``gun shot", ``jackhammer", ``siren", and ``street music".
To ensure correctness during the evaluation phase, the \mbox{UrbanSound8K} dataset was split by its authors into 10 non-overlapping folds \cite{salamon2014us8k} that we used in this work.

On this dataset, we performed zero-shot inference using the \mbox{AudioCLIP} model trained on \mbox{AudioSet}.
Also, the audio encoding head was fine-tuned to the \mbox{UrbanSound8K} dataset in both, standalone and cooperative fashion, and the classification performance in both setups was assessed.

\emph{ESC-50:} The \mbox{ESC-50} dataset provides $2\,000$ single-channel $5\:\si{\second}$ long audio tracks sampled at $44.1\:\si{\kilo \hertz}$.
As the name suggests, the dataset consists of $50$ classes that can be divided into $5$ major groups: \emph{animal}, \emph{natural and water}, \emph{non-speech human}, \emph{interior}, and \emph{exterior} sounds.
To ensure correctness during the evaluation phase, the \mbox{ESC-50} dataset was split by its author into 5 non-overlapping folds \cite{piczak2015esc} that we used in this work.

On this dataset, we performed zero-shot inference using the \mbox{AudioCLIP} model trained on \mbox{AudioSet}.
Also, the audio encoding head was fine-tuned to the \mbox{ESC-50} dataset in both, standalone and cooperative fashion, and the classification performance in both setups was assessed.

\subsection{Data Augmentation} \label{sec:exp_setup:data_aug}
In comparison to the composite CLIP dataset (\autoref{sec:exp_setup:datasets}), the audio datasets provide two orders of magnitude less training samples, which makes overfitting an issue, especially for the \mbox{UrbanSound8K} and \mbox{ESC-50} datasets.
To address this challenge, several data augmentation techniques were applied that we describe in this section.

\emph{Time Scaling:} Simultaneous change of track duration and its pitch is achieved using random scaling along the time axis.
This kind of augmentation combines two computationally expensive ones, namely time stretching and pitch shift.
Being a faster alternative to the combination of the aforementioned techniques, the time scaling in the range of random factors $[-1.5,1.5],\:\sim\mathcal{U}$ provides a lightweight though powerful method to fight overfitting \cite{guzhov2020esrn}.

\emph{Time Inversion:} Inversion of a track along its time axis relates to the random flip of an image, which is an augmentation technique that is widely used in the visual domain.
In this work, random time inversion with the probability of $0.5$ was applied to the training samples similarly to \cite{guzhov2021esrnx}.

\emph{Random Crop and Padding:} Due to the requirement to align track duration before the processing through the model we applied random cropping or padding to the samples that were longer or shorter than the longest track in a non-augmented dataset, respectively.
During the evaluation phase, the random operation was replaced by the center one.

\emph{Random Noise:} The addition of random noise was shown to be helpful to overcome overfitting in visual-realted tasks \cite{hussain2017augment}.
Also, the robustness evaluation of the \mbox{ESResNeXt} model suggested the improved sustainability of the chosen audio encoding model against the additive white Gaussian noise (AWGN) \cite{guzhov2021esrnx}.
In this work, we extended the set of data augmentation techniques using AWGN, whose sound-to-noise ratio varied randomly ($\sim \mathcal{U}$) from $10.0\:\si{\decibel}$ to $120\:\si{\decibel}$.
The probability of the presense of the noise was set to $0.25$.

\subsection{Training} \label{sec:exp_setup:training}
The entire training process was divided into subsequent steps, which made acquisition of the final \mbox{AudioCLIP} model reliable and assured its high performance.
As described in \autoref{sec:model:clip}, we took a \mbox{ResNet}-based \mbox{CLIP} text-image-model pre-trained on its own dataset (\autoref{sec:exp_setup:datasets}) \cite{radford2021clip} and combined it with the \mbox{ESResNeXt} audio-model initialized using \mbox{ImageNet} weights and then pre-trained on the \mbox{AudioSet} dataset \cite{guzhov2021esrnx}.

While the CLIP model was already pre-trained on text-image pairs, we decided to perform an extended \mbox{AudioSet} pre-training of the audio-head first, as it improved performance of the base \mbox{ESResNeXt} model (\autoref{tbl:clf:esrnx}), and then to continue training in a tri-modal setting combining it with two other heads.
Here, the whole \mbox{AudioCLIP} model was trained jointly on the \mbox{AudioSet} dataset using audio snippets, the corresponding video frames and the assigned textual labels.

Finally, audio-head of the trained \mbox{AudioCLIP} model was fine-tuned on the \mbox{UrbanSound8K} and \mbox{ESC-50} datasets in a bimodal manner (audio and text) using sound recordings and the corresponding textual labels.

The trained \mbox{AudioCLIP} model and its audio encoding head were evaluated on the \mbox{ImageNet} dataset as well as on the three audio-datasets: \mbox{AudioSet}, \mbox{UrbanSound8K}, and \mbox{ESC-50}.

\subsubsection{Audio-Head Pre-Training} \label{sec:exp_setup:training:ah_training}
The initialization of the audio-head's parameters was split into two steps.
First, the \mbox{ImageNet}-initialized \mbox{ESResNeXt} model was trained on the \mbox{AudioSet} dataset in a standalone fashion.
Then, the pre-trained audio-head was incorporated into the \mbox{AudioCLIP} model and trained further under the cooperative supervision of the text- and image-heads.

\emph{Standalone:} The first pre-training step implied the use of the \mbox{AudioSet} dataset as a weight initializer.
Here, the \mbox{ESResNeXt} model was trained using the same setup as described in \cite{guzhov2021esrnx}, with the difference in the number of training epochs.
In this work, we increased the training time, which turned out into better evaluation performance on the \mbox{AudioSet} dataset and the subsequent downstream tasks, as described in \autoref{sec:results:clf} and independently quantified.

\emph{Cooperative:} The further pre-training of the audio-head was done jointly with the text- and image-heads.
Here, the pre-trained (in a standalone manner) audio-head was modified slightly through the replacement of its classification layer with a randomly initialized one, whose number of output neurons was the same as the size of \mbox{CLIP's} embedding space.

In this setup, the audio-head was trained as a part of the \mbox{AudioCLIP} model, which made its outputs compatible with the embeddings of the vanilla \mbox{CLIP} model.
Parameters of the two other subnetworks, namely text- and image-head, were frozen during the cooperative pre-training of the audio encoding head, thus, these heads served as teachers in a multi-modal knowledge distillation setup.

The performance of the \mbox{AudioCLIP} model trained in such a fashion was assessed and is described in \autoref{sec:results}.

\subsubsection{AudioCLIP Training} \label{sec:exp_setup:training:full_training}
The joint training of the audio-head made it compatible with the vanilla \mbox{CLIP} model, however, the distribution of images and textual descriptions in the \mbox{AudioSet} dataset does not follow the one from the \mbox{CLIP} dataset.
This could lead to suboptimal performance of the resulting \mbox{AudioCLIP} model on the target dataset as well as on the downstream tasks.

To address this issue, we decided to perform the training of the whole tri-modal model on the \mbox{AudioSet} dataset.
Here, all three modality-dedicated heads were tuned together, making the resulting model take into account the distributions of images and textual descriptions (video frames and names of the assigned \mbox{AudioSet} classes, respectively), in addition to the distribution of audio samples.
The influence of the whole model training on the network's performance in comparison to the audio-head-only training is described in \autoref{sec:results}.

\subsubsection{Audio-Head Fine-Tuning} \label{sec:exp_setup:training:ah_tuning}
The trained \mbox{AudioCLIP} model provides general-purpose multimodal classification, or more general, querying abilities.
However, under some conditions, it is required to acquire a more domain-specific model, which is able to distinguish concepts that differ just slightly.

To address this need, we performed experiments on tuning of the audio encoding head to two target datasets: \mbox{UrbanSound8K} and \mbox{ESC-50}.

\emph{Standalone:} The \mbox{ESResNeXt} model that served as the audio-head demonstrated strong classification abilities on the chosen downstream tasks \cite{guzhov2021esrnx}.
As we performed the \mbox{AudioSet} pre-training step instead of using a pre-trained \mbox{ESResNeXt} model, we fine-tuned it to the \mbox{UrbanSound8K} and \mbox{ESC-50} datasets as well, in order to assess the change of the classification accuracy.

The fine-tuning step was done the same way as in \cite{guzhov2021esrnx}, which implied the replacement of the classification layer with a randomly initialized one, whose number of outputs was defined by the number of targets in the downstream task.
We report the performance of the fine-tuned \mbox{ESResNeXt} model in \autoref{sec:results:clf:ah}.

\emph{Cooperative:} During the fine-tuning of the \mbox{AudioCLIP} model to the downstream tasks, only the parameters of the audio-head were being updated, so the text- and image-heads were frozen at this step.
In comparison to the \mbox{AudioSet} training, which implied a multi-label setup, the corresponding textual class labels from the \mbox{UrbanSound8K} and \mbox{ESC-50} datasets were represented by one class per audio sample.

For the fine-tuned \mbox{AudioCLIP} model, we assess the downstream classification performance as well as the querying performance, as described in \autoref{sec:results:query}.

\subsection{Hyper-Parameters} \label{sec:exp_setup:hparams}
In this work, we trained our model on the \mbox{AudioSet}, \mbox{UrbanSound8K} and \mbox{ESC-50} datasets.
The required hyper-parameters are reported in the current section.

In all training phases, the model parameters were optimized using Stochastic Gradient Descent \cite{polyak1992sgd} optimizer with Nesterov's momentum \cite{nesterov1983momentum} of $0.9$, weight decay of $5 \cdot 10^{-4}$ and batch size of $64$.

The learning rate value decreased exponentially, varying its value $\eta$ and the decrease factor $\gamma$ from $10^{-4}$ and $0.95$, respectively, during the standalone pre-training of the audio-head to $5 \cdot 10^{-5}$ and $0.98$ during the fine-tuning of the \mbox{AudioCLIP} model to the downstream tasks.

The number of epochs was set to $30$ for the \mbox{AudioSet}-based training, and to $50$ for the fine-tuning to the downstream tasks.

\begin{table}[tbp]
\begin{threeparttable}[t]
\caption{Evaluation results of the \mbox{ESResNeXt} model trained on the \mbox{AudioSet} dataset for more epochs. In comparison to the original training, performance improves.}
\label{tbl:clf:esrnx}
\ra{1.0}
\begin{tabularx}{\linewidth}{lXccc}
\toprule
\multicolumn{1}{l}{\multirow{2}{*}{Dataset}} & \multicolumn{1}{c}{\multirow{2}{*}{Score (\%)}} & \multicolumn{3}{c}{ESResNeXt Training} \\
\cmidrule{3-5}
 & & \multicolumn{1}{c}{\cite{guzhov2021esrnx} ($5$ epochs)} & \multicolumn{1}{c}{\qquad\qquad} & \multicolumn{1}{c}{Our ($30$ epochs)} \\
\midrule
\multicolumn{1}{l}{AudioSet} & \multicolumn{1}{c}{mAP} & \multicolumn{1}{c}{28.17} & & \multicolumn{1}{c}{34.14} \\
\addlinespace[0.5em]
\multicolumn{1}{l}{UrbanSound8K} & \multicolumn{1}{c}{\multirow{2}{*}{accuracy}} & \multicolumn{1}{c}{89.14} & & \multicolumn{1}{c}{89.49} \\
\multicolumn{1}{l}{ESC-50} & & \multicolumn{1}{c}{95.20} & & \multicolumn{1}{c}{95.90} \\
\bottomrule
\end{tabularx}
\end{threeparttable}
\end{table}

\subsection{Performance Evaluation} \label{sec:exp_setup:infer}
The model performance was assessed based on two tasks: classification and querying.
While the evaluation of the first was possible for both, audio-head itself and the full \mbox{AudioCLIP} model, the performance on the latter task was assessed for the multimodal network only.

\subsubsection{Classification} \label{sec:exp_setup:infer:clf}
The evaluation of the classification performance was done using the \mbox{AudioCLIP} model as well as its audio-head, namely \mbox{ESResNeXt}.
The latter predicted the class labels directly, as the number of its outputs was equal to the number of targets in the datasets.
For the \mbox{AudioCLIP} model, the classification task implied an intermediate step, which included construction of a target from textual labels \cite{radford2021clip}.

In this work, the performance of the proposed model was evaluated after the training on the \mbox{AudioSet} dataset given audio and/or image as an input.
For the \mbox{UrbanSound8K} and \mbox{ESC-50} datasets, two downstream tasks were evaluated: classification after the training on the target dataset and without the training.
The corresponding accuracies are reported in \autoref{sec:results:clf}.

\subsubsection{Querying} \label{sec:exp_setup:infer:query}

The multimodal nature and symmetry of \mbox{AudioCLIP} allowed to perform querying of samples represented by another modality.
Here, classification can be considered as a sub-task of querying, whose query consists of image and/or audio while the result is represented by text.

In this work, we assessed the querying performance of the trained \mbox{AudioCLIP} model on the \mbox{ImageNet}, \mbox{AudioSet}, \mbox{UrbanSound8K} and \mbox{ESC-50} datasets.
The results include Top-1 Precision/Recall (P@1/R@1) and Mean Average Precision (mAP), and presented in \autoref{sec:results:query}.

\begin{table}[tbp]
\begin{threeparttable}[t]
\caption{Evaluation of \mbox{AudioCLIP} after partial (audio-head) and full training on \mbox{AudioSet}. The latter improves, in general, the results on the downstream tasks.}
\label{tbl:clf:aclp}
\ra{1.0}
\begin{tabularx}{\linewidth}{lXccccccc}
\toprule
\multicolumn{1}{l}{\multirow{2}{*}{Dataset}} & \multicolumn{1}{c}{\multirow{2}{*}{Modality}} & \multicolumn{1}{c}{\multirow{2}{*}{Score (\%)}} & \multicolumn{1}{c}{\quad\quad} & \multicolumn{1}{c}{Training} & \multicolumn{1}{c}{\quad} & \multicolumn{3}{c}{Training} \\
\cmidrule{7-9}
 & & & & \multicolumn{1}{c}{On Target} & & \multicolumn{1}{l}{Audio-Head} & \multicolumn{1}{c}{\qquad} & \multicolumn{1}{l}{Full Model} \\

\midrule

\multicolumn{1}{l}{ImageNet} & \multicolumn{1}{c}{image} & \multicolumn{1}{c}{accuracy} & & & & \multicolumn{1}{c}{40.51} & & \multicolumn{1}{c}{21.79} \\

\addlinespace[0.5em]

\multicolumn{1}{l}{\multirow{3}{*}{AudioSet}} & \multicolumn{1}{c}{image} & \multicolumn{1}{c}{\multirow{3}{*}{mAP}} & & \checkmark & & \multicolumn{1}{c}{\;\;8.93} & & \multicolumn{1}{c}{14.82} \\
 & \multicolumn{1}{c}{audio} & & & \checkmark & & \multicolumn{1}{c}{25.85} & & \multicolumn{1}{c}{28.36} \\
 & \multicolumn{1}{c}{both} & & & \checkmark & & \multicolumn{1}{c}{25.11} & & \multicolumn{1}{c}{32.38} \\

\addlinespace[0.5em]

\multicolumn{1}{l}{\multirow{2}{*}{UrbanSound8K}} & \multicolumn{1}{c}{\multirow{4}{*}{audio}} & \multicolumn{1}{c}{\multirow{4}{*}{accuracy}} & & & & \multicolumn{1}{c}{65.31} & & \multicolumn{1}{c}{68.78} \\
 & & & & \checkmark & & \multicolumn{1}{c}{89.95} & & \multicolumn{1}{c}{\textbf{90.07}} \\

\multicolumn{1}{l}{\multirow{2}{*}{ESC-50}} & & & & & & \multicolumn{1}{c}{69.40} & & \multicolumn{1}{c}{68.60} \\
 & & & & \checkmark & & \multicolumn{1}{c}{96.65} & & \multicolumn{1}{c}{\textbf{97.15}} \\

\bottomrule
\end{tabularx}
\end{threeparttable}
\end{table}

\section{Results} \label{sec:results}

\subsection{Classification} \label{sec:results:clf}

\subsubsection{Audio-Head Only} \label{sec:results:clf:ah}

The extended pre-training ($30$ epochs instead of $5$) on the \mbox{AudioSet} dataset provided an audio encoding head that had increased performance, in comparison to the original training (from $28.17\,\%$ to $34.14\,\%$, mAP).
Such an improvement was also beneficial for the downstream tasks, making the newly trained audio-head to out-perform its base variant on the \mbox{UrbanSound8K} and \mbox{ESC-50} datasets and achieving accuracy of $89.49\,\%$ and $95.90\,\%$, respectively (\autoref{tbl:clf:esrnx}).

\subsubsection{AudioCLIP} \label{sec:results:clf:audioclip}

Our tri-modal training setup -- through the use of video frames -- introduced more diversity into audio-head's target distribution, thus fighting the overfitting issue and further increasing performance in audio classification tasks, in comparison to the audio-only \mbox{ESResNeXt} model.
Also, the joint training of all three heads provided an additional performance boost and the ability to use multiple modalities to perform classification, as well as the zero-shot inference capabilities on previously unseen datasets (\autoref{tbl:clf:aclp}).

\begin{table}[tbp]
\begin{threeparttable}[t]
\caption{Evaluation results on \mbox{UrbanSound8K} (\mbox{US8K}) and \mbox{ESC-50}, accuracy (\%).}
\label{tbl:clf:audio}
\ra{1.0}
\begin{tabularx}{\linewidth}{cXccccccc}

\toprule

 & \multicolumn{1}{l}{\multirow{2}{*}{Model}} & \multicolumn{1}{c}{\multirow{2}{*}{Source}} & \quad\quad & \multicolumn{1}{c}{Training} & \quad\quad & \multicolumn{3}{c}{Target Dataset} \\
 \cmidrule{7-9}
 & & & & \multicolumn{1}{c}{On Target} & & \multicolumn{1}{c}{US8K} & \quad\quad & \multicolumn{1}{c}{ESC-50} \\

\midrule

\parbox[t]{4mm}{\multirow{11}{*}{\rotatebox[origin=c]{90}{\underline{Others}}}} & Human (2015) & \cite{piczak2015esc} & & \;--\; & & \;--\; & & 81.30 \\
 & Piczak-CNN (2015) & \cite{piczak2015cnn} & & \checkmark & & 73.70 & & 64.50 \\
 & SB-CNN (2017) & \cite{salamon2017cnn} & & \checkmark & & 79.00 & & \;--\; \\
 & VGGish + Word2Vec (2019) & \cite{xie2019zero} & & & & \;--\; & & 26.00 \\
 & ESResNet (2020) & \cite{guzhov2020esrn} & & \checkmark & & 85.42 & & 91.50 \\
 & WEANET $N^{4}$ (2020) & \cite{kumar2020weanet} & & \checkmark & & \;--\; & & 94.10 \\
 & DenseNet-201$\,\times\,5$, ensemble (2020) & \cite{palanisamy2020densenet} & & \checkmark & & 87.42 & & 92.89 \\
 & VGGish + Word2Vec + GloVe (2021) & \cite{xie2021zero} & & & & \;--\; & & 33.00 \\
 & ESResNeXt (2021) & \cite{guzhov2021esrnx} & & \checkmark & & 89.14 & & 95.20 \\
 & AST (2021) & \cite{gong2021ast} & & \checkmark & & \;--\; & & 95.60 \\
 & ERANN (2021) & \cite{verbitskiy2021erann} & & \checkmark & & \;--\; & & 96.10 \\

\cmidrule{2-9}

\parbox[t]{4mm}{\multirow{5}{*}{\rotatebox[origin=c]{90}{\underline{Ours}}}} & Audio-Head (ESResNeXt, our training) & & & \checkmark & & 89.49 & & 95.90 \\

\addlinespace[0.5em]

 & \multicolumn{1}{l}{\multirow{2}{*}{AudioCLIP (partial training)}} & & & & & 65.31 & & \textbf{69.40} \\
 & & & & \checkmark & & 89.95 & & 96.65 \\
 & \multicolumn{1}{l}{\multirow{2}{*}{AudioCLIP (full training)}} & & & & & \textbf{68.78} & & 68.60 \\
 & & & & \checkmark & & \textbf{90.07} & & \textbf{97.15} \\

\bottomrule
\end{tabularx}
\end{threeparttable}
\end{table}

\emph{Partial Training:} The training of the audio-head under the supervision of the text- and image-heads already allows to out-perform current state-of-the-art results on the \mbox{UrbanSound8K} and \mbox{ESC-50} datasets by achieving accuracy of $89.95\,\%$ and $96.65\,\%$, respectively.

Moreover, even the partial training of the \mbox{AudioCLIP} model sets new highest zero-shot classification accuracy on the \mbox{ESC-50} dataset ($69.40\,\%$, \autoref{tbl:clf:audio}) as well as out-performs performance of the commonly trained baseline CNN ($64.50\,\%$, \autoref{tbl:clf:audio}).

\emph{Full Training:} The joint training of the \mbox{AudioCLIP} model provides further performance improvements in comparison to the partial one.
Such a trained \mbox{AudioCLIP} model sets the new state-of-the-art classification accuracy on the \mbox{UrbanSound8K} and \mbox{ESC-50} datasets ($90.07\,\%$ and $97.15\,\%$, respectively).
Also, given the full model training setup, a new zero-shot classification baseline was set for the \mbox{UrbanSound8K} dataset ($68.78\,\%$, \autoref{tbl:clf:audio}).

\subsection{Querying} \label{sec:results:query}
The original \mbox{CLIP} model introduced the ability to perform querying using both supported modalities -- text and image -- in any direction.
Given a query (e.g., text), model provided similarity scores for the samples represented by another (visual) modality.
Thus, given a dataset and a modality, the set of queries was defined by the unique samples of the chosen modality.
In this work, we added the support of audio, enabling the model to query between text, images and audio in any combination.
We evaluated the querying performance on the \mbox{ImageNet}, \mbox{AudioSet}, \mbox{UrbanSound8K} and \mbox{ESC-50} datasets and summarized it in \autoref{tbl:query}.

\begin{table}[tbp]
\begin{threeparttable}[t]
\caption{Querying scores of \mbox{AudioCLIP} after partial and full training on \mbox{AudioSet}. The latter in general improves results on \mbox{AudioSet} and the downstream tasks.}
\label{tbl:query}
\ra{1.0}
\begin{tabularx}{\linewidth}{Xcccccccccc}

\toprule

\multicolumn{1}{l}{\multirow{2}{*}{Dataset}} & \multicolumn{2}{c}{Modality} & & \multicolumn{3}{c}{Audio-Head} & & \multicolumn{3}{c}{Full Model} \\
 
 \cmidrule{2-3}
 \cmidrule{5-7}
 \cmidrule{9-11}
 
 & \multicolumn{1}{c}{Query} & \multicolumn{1}{c}{Result} & & \multicolumn{1}{c}{P@1} & \multicolumn{1}{c}{R@1} & \multicolumn{1}{c}{mAP} & & \multicolumn{1}{c}{P@1} & \multicolumn{1}{c}{R@1} & \multicolumn{1}{c}{mAP} \\

\midrule

\multicolumn{1}{l}{ImageNet} & \multicolumn{1}{c}{text} & \multicolumn{1}{c}{image} & & \multicolumn{1}{c}{\;\;5.42} & \multicolumn{1}{c}{84.15} & \multicolumn{1}{c}{\textbf{52.91}} &  & \multicolumn{1}{c}{\;\;1.61} & \multicolumn{1}{c}{89.00} & \multicolumn{1}{c}{33.13} \\

\addlinespace[0.5em]

\multicolumn{1}{l}{\multirow{4}{*}{AudioSet}} & \multicolumn{1}{c}{text} & \multicolumn{1}{c}{image} & & \multicolumn{1}{c}{\;\;0.81} & \multicolumn{1}{c}{46.79} & \multicolumn{1}{c}{\;\;9.51} &  & \multicolumn{1}{c}{\;\;1.31} & \multicolumn{1}{c}{74.95} & \multicolumn{1}{c}{\textbf{17.22}} \\
 & \multicolumn{1}{c}{text} & \multicolumn{1}{c}{audio} & & \multicolumn{1}{c}{\;\;2.51} & \multicolumn{1}{c}{84.38} & \multicolumn{1}{c}{23.54} &  & \multicolumn{1}{c}{\;\;5.33} & \multicolumn{1}{c}{76.13} & \multicolumn{1}{c}{\textbf{30.79}} \\
 & \multicolumn{1}{c}{audio} & \multicolumn{1}{c}{image} & & \multicolumn{1}{c}{\;\;0.62} & \multicolumn{1}{c}{56.39} & \multicolumn{1}{c}{\;\;5.45} &  & \multicolumn{1}{c}{\;\;1.03} & \multicolumn{1}{c}{52.12} & \multicolumn{1}{c}{\textbf{\;\;7.22}} \\
 & \multicolumn{1}{c}{image} & \multicolumn{1}{c}{audio} & & \multicolumn{1}{c}{\;\;0.61} & \multicolumn{1}{c}{61.94} & \multicolumn{1}{c}{\;\;4.86} &  & \multicolumn{1}{c}{\;\;1.20} & \multicolumn{1}{c}{53.15} & \multicolumn{1}{c}{\textbf{\;\;6.86}} \\

\addlinespace[0.5em]

\multicolumn{1}{l}{UrbanSound8K} & \multicolumn{1}{c}{\multirow{2}{*}{text}} & \multicolumn{1}{c}{\multirow{2}{*}{audio}} & & \multicolumn{1}{c}{40.81} & \multicolumn{1}{c}{47.69} & \multicolumn{1}{c}{77.43} &  & \multicolumn{1}{c}{42.28} & \multicolumn{1}{c}{48.18} & \multicolumn{1}{c}{\textbf{80.04}} \\

\multicolumn{1}{l}{ESC-50} & & & & \multicolumn{1}{c}{51.25} & \multicolumn{1}{c}{85.20} & \multicolumn{1}{c}{\textbf{77.20}} &  & \multicolumn{1}{c}{51.78} & \multicolumn{1}{c}{80.40} & \multicolumn{1}{c}{76.97} \\

\bottomrule

\end{tabularx}
\end{threeparttable}
\end{table}

\emph{Image by Text:} In this setup, all unique sets of class names assigned to the samples from a target dataset were collected and served as textual queries while the results were represented by images (\mbox{ImageNet}, \mbox{AudioSet}).
Thus, only the visual samples possessing the same set of labels were considered as relevant results.

For the \mbox{AudioSet} dataset, the full training contributed to the increase of the performance score measured by mAP.
However, such a training led to the decrease of the querying performance on the \mbox{ImageNet} dataset, as its distribution is likely different from the \mbox{AudioSet} one.

\emph{Audio by Text:} Having the same type of query as in the previous setup, here, the result was represented by an audio recording and considered correct if the labels matched the query.

On the \mbox{AudioSet} and \mbox{UrbanSound8K} datasets, the full training increases the querying performance.
For the \mbox{ESC-50} dataset it is not the case, however, the gap is not large and is close to be marginal.

\emph{Audio by Image and Vice Versa:} For both types of queries -- audio by image and image by audio -- the full training of the \mbox{AudioCLIP} model was beneficial in terms of querying performance (mAP).

\section{Conclusion} \label{sec:conclusion}
In this work, we extended the \mbox{CLIP} model \cite{radford2021clip} from textual and visual modalities to audio using an effective sound classification model \cite{guzhov2021esrnx}.

The proposed \mbox{AudioCLIP} model achieves new state-of-the-art classification results on two datasets: \mbox{UrbanSound8K} ($90.07\,\%$) and \mbox{ESC-50} ($97.15\,\%$).
\ifreview
\else
To ease reproducibility, the details on hyper-parameters and implementation as well as weights of the trained models are made available for the community\footnote[1]{\url{https://github.com/AndreyGuzhov/AudioCLIP}}.
\fi

Additionally, for the zero-shot inference, our model out-performs previous approaches on the \mbox{ESC-50} dataset with a large gap ($69.40\,\%$) and sets a baseline for the \mbox{UrbanSound8K} dataset ($68.78\,\%$).

We also evaluated the performance of our model on cross-modal querying tasks as well as the influence of the partial and full training on the results in classification and querying tasks.

In the future, we would like to further investigate the performance of the proposed \mbox{AudioCLIP} model on a wider variety of datasets and tasks.
Also, changing the backbones of image- and audio-heads to more powerful networks could further improve the model performance.

%
%
\bibliographystyle{splncs04}
\bibliography{egbib}

\begin{thebibliography}{10}
\providecommand{\url}[1]{\texttt{#1}}
\providecommand{\urlprefix}{URL }
\providecommand{\doi}[1]{https://doi.org/#1}

\bibitem{akbari2021vatt}
Akbari, H., Yuan, L., Qian, R., Chuang, W.H., Chang, S.F., Cui, Y., Gong, B.:
  Vatt: Transformers for multimodal self-supervised learning from raw video,
  audio and text. arXiv preprint arXiv:2104.11178  (2021)

\bibitem{alayrac2020versatile}
Alayrac, J.B., Recasens, A., Schneider, R., Arandjelovi{\'c}, R., Ramapuram,
  J., De~Fauw, J., Smaira, L., Dieleman, S., Zisserman, A.: Self-supervised
  multimodal versatile networks. arXiv preprint arXiv:2006.16228  (2020)

\bibitem{chollet2017resnext}
Chollet, F.: Xception: Deep learning with depthwise separable convolutions. In:
  Proceedings of the IEEE conference on computer vision and pattern
  recognition. pp. 1251--1258 (2017)

\bibitem{deng2009imagenet}
Deng, J., Dong, W., Socher, R., Li, L.J., Li, K., Fei-Fei, L.: Imagenet: A
  large-scale hierarchical image database. In: 2009 IEEE conference on computer
  vision and pattern recognition. pp. 248--255. Ieee (2009)

\bibitem{dosovitskiy2020vit}
Dosovitskiy, A., Beyer, L., Kolesnikov, A., Weissenborn, D., Zhai, X.,
  Unterthiner, T., Dehghani, M., Minderer, M., Heigold, G., Gelly, S., et~al.:
  An image is worth 16x16 words: Transformers for image recognition at scale.
  arXiv preprint arXiv:2010.11929  (2020)

\bibitem{dzabraev2021mdmmt}
Dzabraev, M., Kalashnikov, M., Komkov, S., Petiushko, A.: Mdmmt: Multidomain
  multimodal transformer for video retrieval. In: Proceedings of the IEEE/CVF
  Conference on Computer Vision and Pattern Recognition. pp. 3354--3363 (2021)

\bibitem{gemmeke2017audioset}
Gemmeke, J.F., Ellis, D.P., Freedman, D., Jansen, A., Lawrence, W., Moore,
  R.C., Plakal, M., Ritter, M.: Audio set: An ontology and human-labeled
  dataset for audio events. In: 2017 IEEE International Conference on
  Acoustics, Speech and Signal Processing (ICASSP). pp. 776--780. IEEE (2017)

\bibitem{gong2021ast}
Gong, Y., Chung, Y.A., Glass, J.: Ast: Audio spectrogram transformer (2021)

\bibitem{guzhov2020esrn}
Guzhov, A., Raue, F., Hees, J., Dengel, A.: Esresnet: Environmental sound
  classification based on visual domain models. In: 25th International
  Conference on Pattern Recognition (ICPR). pp. 4933--4940 (January 2021)

\bibitem{guzhov2021esrnx}
Guzhov, A., Raue, F., Hees, J., Dengel, A.: Esresne(x)t-fbsp: Learning robust
  time-frequency transformation of audio. In: 2021 International Joint
  Conference on Neural Networks (IJCNN) (2021)

\bibitem{he2016resnet}
He, K., Zhang, X., Ren, S., Sun, J.: Deep residual learning for image
  recognition. In: Proceedings of the IEEE conference on computer vision and
  pattern recognition. pp. 770--778 (2016)

\bibitem{hussain2017augment}
Hussain, Z., Gimenez, F., Yi, D., Rubin, D.: Differential data augmentation
  techniques for medical imaging classification tasks. In: AMIA Annual
  Symposium Proceedings. vol.~2017, p.~979. American Medical Informatics
  Association (2017)

\bibitem{islam2019soundsemantics}
Islam, M.T., Nirjon, S.: Soundsemantics: exploiting semantic knowledge in text
  for embedded acoustic event classification. In: Proceedings of the 18th
  International Conference on Information Processing in Sensor Networks. pp.
  217--228 (2019)

\bibitem{kim2019audiocaps}
Kim, C.D., Kim, B., Lee, H., Kim, G.: Audiocaps: Generating captions for audios
  in the wild. In: Proceedings of the 2019 Conference of the North American
  Chapter of the Association for Computational Linguistics: Human Language
  Technologies, Volume 1 (Long and Short Papers). pp. 119--132 (2019)

\bibitem{kumar2020weanet}
Kumar, A., Ithapu, V.: A sequential self teaching approach for improving
  generalization in sound event recognition. In: International Conference on
  Machine Learning. pp. 5447--5457. PMLR (2020)

\bibitem{nesterov1983momentum}
Nesterov, Y.: A method of solving a convex programming problem with convergence
  rate o (1/k\^{} 2) o (1/k2). In: Sov. Math. Dokl. vol.~27 (1983)

\bibitem{palanisamy2020densenet}
Palanisamy, K., Singhania, D., Yao, A.: Rethinking cnn models for audio
  classification (2020)

\bibitem{piczak2015cnn}
Piczak, K.J.: Environmental sound classification with convolutional neural
  networks. In: 2015 IEEE 25th International Workshop on Machine Learning for
  Signal Processing (MLSP). pp.~1--6. IEEE (2015)

\bibitem{piczak2015esc}
Piczak, K.J.: Esc: Dataset for environmental sound classification. In:
  Proceedings of the 23rd ACM international conference on Multimedia. pp.
  1015--1018 (2015)

\bibitem{polyak1992sgd}
Polyak, B.T., Juditsky, A.B.: Acceleration of stochastic approximation by
  averaging. SIAM journal on control and optimization  \textbf{30}(4),
  838--855 (1992)

\bibitem{radford2021clip}
Radford, A., Kim, J.W., Hallacy, C., Ramesh, A., Goh, G., Agarwal, S., Sastry,
  G., Askell, A., Mishkin, P., Clark, J., Krueger, G., Sutskever, I.: Learning
  transferable visual models from natural language supervision (2021)

\bibitem{radford2019transformer}
Radford, A., Wu, J., Child, R., Luan, D., Amodei, D., Sutskever, I.: Language
  models are unsupervised multitask learners. OpenAI blog  \textbf{1}(8), ~9
  (2019)

\bibitem{sailor2017convrbm}
Sailor, H.B., Agrawal, D.M., Patil, H.A.: Unsupervised filterbank learning
  using convolutional restricted boltzmann machine for environmental sound
  classification. In: INTERSPEECH. pp. 3107--3111 (2017)

\bibitem{salamon2017cnn}
Salamon, J., Bello, J.P.: Deep convolutional neural networks and data
  augmentation for environmental sound classification. IEEE Signal Processing
  Letters  \textbf{24}(3),  279--283 (2017)

\bibitem{salamon2014us8k}
Salamon, J., Jacoby, C., Bello, J.P.: A dataset and taxonomy for urban sound
  research. In: Proceedings of the 22nd ACM international conference on
  Multimedia. pp. 1041--1044 (2014)

\bibitem{teolis1998fbsp}
Teolis, A., Benedetto, J.J.: Computational signal processing with wavelets,
  vol.~182. Springer (1998)

\bibitem{tokozume2017envnet}
{Tokozume}, Y., {Harada}, T.: Learning environmental sounds with end-to-end
  convolutional neural network. In: 2017 IEEE International Conference on
  Acoustics, Speech and Signal Processing (ICASSP). pp. 2721--2725 (March
  2017). \doi{10.1109/ICASSP.2017.7952651}

\bibitem{tokozume2017envnetv2}
Tokozume, Y., Ushiku, Y., Harada, T.: Learning from between-class examples for
  deep sound recognition (2017), \url{https://arxiv.org/abs/1711.10282}

\bibitem{vaswani2017attention}
Vaswani, A., Shazeer, N., Parmar, N., Uszkoreit, J., Jones, L., Gomez, A.N.,
  Kaiser, L., Polosukhin, I.: Attention is all you need. arXiv preprint
  arXiv:1706.03762  (2017)

\bibitem{verbitskiy2021erann}
Verbitskiy, S., Vyshegorodtsev, V.: Eranns: Efficient residual audio neural
  networks for audio pattern recognition (2021)

\bibitem{wang2021multimodal}
Wang, L., Luc, P., Recasens, A., Alayrac, J.B., Oord, A.v.d.: Multimodal
  self-supervised learning of general audio representations. arXiv preprint
  arXiv:2104.12807  (2021)

\bibitem{xie2019zero}
Xie, H., Virtanen, T.: Zero-shot audio classification based on class label
  embeddings. In: 2019 IEEE Workshop on Applications of Signal Processing to
  Audio and Acoustics (WASPAA). pp. 264--267. IEEE (2019)

\bibitem{xie2021zero}
Xie, H., Virtanen, T.: Zero-shot audio classification via semantic embeddings.
  IEEE/ACM Transactions on Audio, Speech, and Language Processing  \textbf{29},
   1233--1242 (2021)

\bibitem{zhang2021sscl}
Zhang, J., Xu, X., Shen, F., Lu, H., Liu, X., Shen, H.T.: Enhancing
  audio-visual association with self-supervised curriculum learning. In:
  Proceedings of the AAAI Conference on Artificial Intelligence. vol.~35, pp.
  3351--3359 (2021)

\end{thebibliography}

\end{document}